\documentclass[twocolumn,twocolappendix,tighten]{aastex7}

\usepackage{amsmath}
\usepackage{enumitem}
\usepackage{gensymb}
\usepackage{natbib}
\usepackage{xspace}

\AtBeginEnvironment{thebibliography}{\linespread{1}\selectfont}

\def\m87{M87$^*$\xspace}
\defcitealias{M87Paper1}{\m87~Paper~I}
\defcitealias{2024A&A...681A..79E}{\m87~2018}
\defcitealias{2021ApJ...910L..12E}{\m87~Paper~VII}
\defcitealias{2019ApJ...875L...6E}{\m87~Paper~VI}
\defcitealias{M87_paper4}{\m87~Paper IV}
\defcitealias{2021ApJ...910L..13E}{\m87~Paper~VIII}
\defcitealias{2019ApJ...875L...5E}{\m87~Paper~V}

\begin{document}	

\title{New estimates of the spin and accretion rate of the black hole M87$^*$}

\author[0009-0006-0378-3647]{Michael Drew}
\affiliation{Jeremiah Horrocks Institute, University of Central Lancashire, Preston PR1 2HE, UK}
\email{MDrew1@uclan.ac.uk}

\author[0009-0003-7659-4642]{Joshua S. Stanway}
\affiliation{Jeremiah Horrocks Institute, University of Central Lancashire, Preston PR1 2HE, UK}
\email{JSStanway@uclan.ac.uk}

\author[0000-0002-5605-4660]{Brett A. Patterson}
\affiliation{Jeremiah Horrocks Institute, University of Central Lancashire, Preston PR1 2HE, UK}
\email{BAPatteson@uclan.ac.uk}

\author[0000-0001-5103-4591]{Timothy J. Walton}
\affiliation{Jeremiah Horrocks Institute, University of Central Lancashire, Preston PR1 2HE, UK}
\email{TWalton@uclan.ac.uk}

\author[0000-0003-1140-2761]{Derek Ward-Thompson}
\affiliation{Jeremiah Horrocks Institute, University of Central Lancashire, Preston PR1 2HE, UK}
\email{DWard-Thompson@uclan.ac.uk}

\begin{abstract}
	In this paper we use the imaging results of M87* from the EHT to calculate the rotational velocity of the inner edge of the accretion disk and find a value of  $\sim$0.14c. We then calculate the dimensionless spin parameter, $a$, of the black hole, obtaining a value of $a \sim 0.8$. We deduce that this is probably a lower limit. We go on to use the results of the EHT polarization study of the magnetic field direction in the accretion disk of M87* as a proxy for the direction of motion of the spiralling accreting matter in this highly ionized disk. This direction is defined by the vector sum of the tangential rotation velocity and the inward radial accretion velocity. We thus calculate the accretion velocity to be $\sim$ (7 $\pm$ {0.7}) $\times$ $10^7$~ms$^{-1}$. We go on to estimate a range of values for the accretion rate from the inner disk to be $\sim$4 $\times$ 10$^{{-5}}$ to $\sim$4 $\times$ 10$^{-1}$ M$_{\odot}$yr$^{-1}$, and a range of values for the accretion power to be $\sim$10$^{34}$ to 10$^{38}$ J/s. This is the same range as the power of the jet, making it consistent with accretion-driven jet models.
\end{abstract}


\section{Introduction}\label{sec:intro}
The study of supermassive black holes (SMBHs) is an important topic in astrophysics, given observational evidence that they exist at the centre of nearly all local large galaxies \citep{Berti_2008} and play a significant role in the formation and evolution of their host galaxies -- e.g.: \cite{richstone1998supermassive}; \cite{reynolds2019observing}.  Given that measuring the spin of supermassive black holes can contribute to our understanding of astrophysical processes at the core of galaxies, it has become an important area of research. However, although angular momentum is a fundamental property of black holes, along with mass and electric charge, it has proved challenging to constrain spin parameters \citep{2013mams.book.....B}. 

Observational studies of M87* have exploited advances in very-long-baseline interferometry (VLBI). In particular, transformational progress by the Event Horizon Telescope
Collaboration (EHTC) has achieved an angular resolution comparable to the event horizon of M87* \citep[known as \citetalias{M87Paper1}]{M87Paper1}.  The image produced from these observations (taken in 2017)
produced world-wide interest (see Figure~1, image).

The recent publication of the second epoch, in 2018, of EHT observations
of M87* \citep[known as \citetalias{2024A&A...681A..79E}, see Figure~1, contours]{2024A&A...681A..79E} has revealed some similarities to, and differences from, the first epoch in 2017.  

Figure~1 shows the 2017 epoch image with the 2018 data overlaid as contours. The most striking similarity is with regard to the position of the brightest part of the ring. As the crosses show, the brightest part of the ring is consistent, within a quarter of a beam-width (the sizes of the crosses are not error-bars), with not having moved. However, what has changed is the extent of the bright region, which has decreased between 2017 and 2018.

We here hypothesise that the brightest spot that remains static is actually due to the relativistic rotation of the disk, whereby the enhanced brightness is caused by relativistic beaming.
The remainder of the bright region that fades between 2017 and 2018 could be caused by variations in the position and relative brightness of the base of the jet.

New techniques have been developed to overcome the current limitations on constraining the spin of M87* by analysis of the black hole shadow. The approach taken in this paper is to use the relativistic Doppler beaming effect, visible in the images of M87*, to derive an estimate for the rotation velocity of the disk. We also consider the measured polarization angles \citep[hereafter \citetalias{2021ApJ...910L..12E}]{2021ApJ...910L..12E} of the disk around M87* to estimate the accretion velocity and accretion rate.\\\\

\begin{figure} 
	\centering
	\includegraphics[width=0.9\columnwidth]{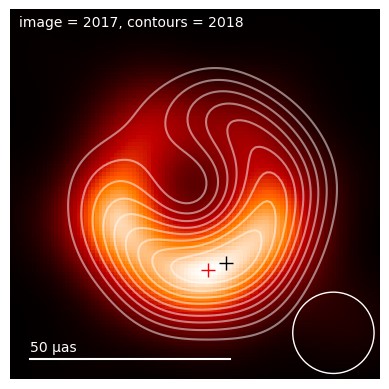}
	\caption{M87* from the observing programs in 2017 (image) and 2018 (contours).  The location of peak brightness in 2017 (red cross) and 2018 (black cross) is consistent to within a quarter of one beam (white circle), although the extent of the bright region has decreased from 2017 to 2018.}
\end{figure}

\section{Methods and Results}\label{sec:methods}
Given the estimated age of M87* of over 10 Gyr, it is assumed that the angular momentum of the inner accretion disk matches the spin of the black hole, as is expected to develop over timescales of only 10 Myr \citep{Thorne_1974}. The velocity of rotating matter is estimated using observational data  and interpreting it as relativistic Doppler beaming. This has been shown to be a valid calculation, even when the disk is close to edge-on, as in this case \citep{Medeiros_2022}. Estimating the radius of the inner edge of the accretion disk then leads to an estimate of the spin parameter.

This method has a number of caveats, including that it is based on the assumption that the emission seen is all coming from the inner accretion disk, and that there is no significant amount of matter between this and the event horizon. It also assumes that the asymmetry in brightness is wholly due  to relativistic beaming, the SED is approximately flat near the EHT bands and the effects of frame-dragging on the photon trajectory can be neglected.

\subsection{The rotation of the accretion disk}
As the accretion disk spins rapidly, relativistic Doppler beaming affects the apparent brightness of the ring. On one side of the disk, the fast-rotating plasma approaches us and, on the other side, it recedes.  

The Doppler factor measures the strength of the relativistic Doppler beaming, taking into account the angle between the direction of motion of the matter and the line of sight \citep{Narayan_2005}.  In this case, we use the equation for the Doppler factor to estimate the velocity \citep{1982PNAS...79.1340J}, 
by comparing the difference in brightness of the approaching $(D_\text{app})$ and receding $(D_\text{rec})$ matter \citep{1986rpa..book.....R}, as follows:
\begin{eqnarray}
    D_\text{\rm app} &=& \left[\gamma\left(1 - \frac{v\cos\theta}{c}\right)\right]^{-1} \, ,\\
    D_\text{\rm rec} &=& \left[\gamma\left(1 + \frac{v\cos\theta}{c}\right)\right]^{-1} \, ,
\end{eqnarray}
\noindent
where $v$ is the velocity of the matter, $\gamma$ is the Lorentz factor,  which is given by $(1-v^2/c^2)^{-1/2}$, and the angle of inclination between the direction of motion of the matter and the observer's line of sight, $\theta$ is approximately $17^{\degree}$ \citep[known as \citetalias{2019ApJ...875L...6E}]{2019ApJ...875L...6E}. The Lorentz invariant for relativistic Doppler beaming is ($I_\nu /\nu^3$) 
\citep{1986rpa..book.....R}. Thus the equations for the Doppler factor, $D$, become:
\begin{eqnarray}
	D_{\rm app} &=&
	\left[ \gamma \left( 1 - \frac{0.96V_{R}}{c} \right) \right]^{-1} \, ;\\
	D_{\rm rec} &=&
	\left[\gamma\left(1 + \frac{0.96V_{R}}{c}\right)\right]^{-1} \, .
\end{eqnarray}
Figure~2 shows a cross-cut diametrically through the brightest part of the ring
\citep[hereafter \citetalias{M87_paper4}]{M87_paper4} for each of the three main image reconstruction methods used by the EHT Collaboration. The two peaks in each plot show the two sides of the ring, the bright part and the faint part. In each case we took the ratio of the maxima of the two peaks, as well as the ratio of the areas under the curves out to the full-width at half maximum (FWHM). The former gave a ratio of 2.1 $\pm$ 0.1 and the latter yielded a value of 2.4 $\pm$ 0.2. The weighted mean of these two ratios is therefore 2.2 $\pm$ 0.3. This is the value that we adopt. Putting this all together gives us
$(D_\text{app}/D_\text{rec})^3 \, \sim \, 2.25$ -- see also: \citet{Misner:1973prb}. Therefore
\begin{equation*}
\left[\gamma\left(1 - \frac{0.96V_{R}}{c}\right)\right]^{-3} = 2.25 \left[\gamma\left(1 + 
\frac{0.96V_{R}}{c}\right)\right]^{-3},
\end{equation*}
\noindent 
where $V_R$ is the velocity at radius $R$, at the inner edge of the accretion disk. Thus, we estimate that $V_{R}$ is approximately 0.14c, or  $(4.2\, \pm \, 0.4) \, \times \, 10^7$ m~s$^{-1}$.

\begin{figure} 
	\centering
	\includegraphics[width=1.0\columnwidth]{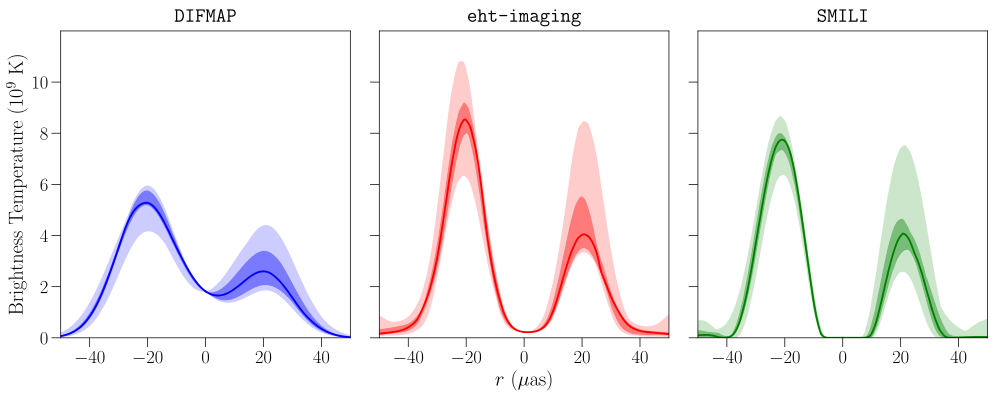}
	\caption{A cross section of brightness vs relative position of each side of the ring (in $\mu$arcsec) for each of the three main EHT image reconstruction methods, as applied to the 2017 data.
    Image credit: \citetalias{M87_paper4} fig. 37.}
\end{figure}

\subsection{Estimating the spin}
There is expected to be differential rotation across the disk, as the speed of rotation decreases with increasing radius.  We make the practical assumption that the inner edge of the accretion disk, is at the innermost stable circular orbit (ISCO), with radius $R_{\rm ISCO}$ \citep{Reynolds_2021}. Some models predict emission from further in, but this is not confirmed by other models, so we ignore this possibility {(but we note that this is an assumption of our method).

Furthermore, due to the near face-on viewing angle, we assume that the lensing effect of the Kerr spacetime around M87* on the overall image size is negligible, and so we assume that $R_{\rm ISCO}$ can be measured directly from images such as Figures 1 and 2. Our best estimate of the value of this inner radius is 20 $\pm$ 2 $\mu$-arcsec. This is consistent with previous estimates, which use a variety of techniques -- e.g.: \cite{2012Sci...338..355D}; \cite{article}. Assuming a distance to M87 of 16.8 Mpc, this leads to an estimate of $R_{\rm ISCO}$ to be roughly (5.5 $\pm$ 0.5) $\times$ 10$^{13}$~m. This is $\sim$3 Schwarzchild radii, which is also consistent with ISCOs allowed range for a Kerr black hole \citep{Bardeen_1972}.

We now assume that angular momentum is conserved inside $R_{\rm ISCO}$. This assumption is based on there being much less material inside $R_{\rm ISCO}$, and hence viscosity can be ignored in this region.
Some GRMHD models predict material inside $R_{\rm ISCO}$ but not all, so we take the very low density
inside this radius as an assumption, with this caveat. Therefore, we take the product of $V \times R$ per unit mass to be constant from $R_{\rm ISCO}$ down to the black hole event horizon. This would give a value for the black hole angular momentum, $J$, of
\begin{equation}   
J \, = \, M_{BH} \, V_{R} \, R_{\rm ISCO} \, .
\end{equation}
\noindent
The dimensionless spin parameter, $a$, is defined as
\begin{equation}   
a \, = \, \frac{J  \, c}{G \, M_{BH}^2} \, 
= \, \frac{V_R \, R_{\rm ISCO} \, c}{G \, M_{BH}} \, .
\end{equation}
\noindent
If we use $M_{BH} \, = \, (6.5 \, \pm \, 0.1) \, \times \, 10^{9} \, M_{\odot}$, \citep{M87Paper1}, then we find $a \, \sim \, 0.8 \,.$

We note that our method automatically assumes a Minkowski metric. In Appendix A we
make the corrections for a Schwarzchild and a Kerr metric. Both cases tend to raise the value of $a$ to nearer its upper limit of 0.998. However, in order to estimate the Kerr spin we have to start from a value of 0.8. Nevertheless, our value of 0.8 would appear to be a lower limit. Previous estimates span the range of $\sim$~0.1 to 0.98, so our method appears to rule out the lower part of this range.

\subsection{The accreting matter} 
Matter is accreting onto the black hole M87*, and this occurs through the accretion disk, in which the matter is highly ionised. Therefore, we make the plausible assumption that the matter is tied to the magnetic field in the disk. The polarisation study of M87* \citepalias{2021ApJ...910L..12E} showed that the magnetic field in the disk is neither radial nor tangential but rather, at each point, lies at an angle to  the tangent vector at that point \citepalias{2021ApJ...910L..12E}. We made measurements of these data to estimate this angle. An average was made for all estimates in all positions over all four days of the observations.  We label this angle $\alpha$ and find it to be (58.5 $\pm$ 0.6)$^{\degree}$,
assuming that the magnetic field is orthogonal to the measured polarisation. We ignore the effects of special and general relativity on the photon trajectories, which will affect the polarisation pattern, as mentioned by \cite[hereafter \citetalias{2021ApJ...910L..13E}]{2021ApJ...910L..13E}.

Previous authors have all assumed the accretion disk is 100\% ionized, either implicitly or otherwise \citep{Russell_2015}. We follow the same assumption. Therefore, we assume that the matter is tied to the magnetic field and spiralling inward in the disk at the polarization angle, $\alpha$, with 
total velocity, $V_{T}$, which is a combination of the rotational velocity, $V_{R}$, and the accretion velocity, $V_{A}$. Simple geometry shows us that $V_A \, = \, V_R$ tan $\alpha$. Hence $V_A$, the component of the plasma velocity heading towards the black hole, is found to be 7 $\times$ $10^7$~ms$^{-1}$.

The error-bar on this, introduced by the error on $\alpha$, and including all of the above errors, gives
a value of V$_A$ $=$ (7 $\pm$ 0.7) $\times$ $10^7$~ms$^{-1}$, or (0.23 $\pm$ 0.02)c. Given this estimate of the accretion velocity, we can estimate the accretion rate.

We start by envisaging the accretion rate across part of an imaginary sphere of radius $R_{\rm ISCO}$, where there is only accretion across a fraction of the area of the sphere either side of the equatorial plane of the disk -- see Figure~B1.

We first use the `thick disk' approximation and estimate the angle subtended by the inner edge of the disk to be a maximum of $\approx$ $\pm$ 30$^{\degree}$ above and below the mid-plane of the disk, or $\approx$~1.0 radian in total, but acknowledge that this is extremely uncertain. This leads us to the equation for the accretion rate (see Appendix B):

\begin{equation}
	\dot M \, = \, (2\pi) \, V_A \, R_{\rm ISCO}^2 \, \rho \, ,
\end{equation}

\noindent
where the material has accretion velocity $V_A$, and density $\rho$, at radius $R_{\rm ISCO}$. 

If we instead use a `thin disk' approximation of $\approx$~$\pm$~3$^{\degree}$ above and below the mid-plane of the disk, or $\approx$~0.1 radians in total, the term in brackets becomes $(\pi^2/15)$ and the estimate of the accretion rate decreases by a factor of $\sim$~10 (see Appendix for details of this calculation).

The upper limit to the emission region plasma density, $\rho$, is estimated as equivalent to 
$n(e)$ being  $\sim~10^7$~cm$^{-3}$ \citepalias{2021ApJ...910L..13E}. Assuming the gas is completely ionised and assuming equal numbers of electrons and protons (whilst ignoring the mass of the electrons),
the upper limit to the density, $\rho$, can therefore be approximated as $1.7 \, \times \, 10^{-14}$ kg m$^{-3}$. We acknowledge the extreme uncertainty in this value, and take it as an upper limit.
Some estimates of the density are up to three orders of magnitude lower. 

Using the above value for the accretion velocity, $V_A$, the higher estimated density, and the `thick disk' approximation, we obtain an upper limit for the accretion rate, $\dot{M}$, of
$\sim$ 2.3 $\times$ 10$^{22}$ kg~s$^{-1}$, or $\sim$~0.4 $M_{\odot}$~yr$^{-1}$. 

The lower limit, based on the density being potentially three orders of magnitude lower,
and the accretion rate being potentially an order of magnitude lower (the `thin disk' approximation),
is therefore $\sim$~4 $\times$ 10$^{-5}$ $M_{\odot}$~yr$^{-1}$. 

\section{Discussion and Conclusions} \label{sec:Discussion}
Our mass accretion rate estimate is in the range $\sim$~4 $\times$ 10$^{-5}$ to $\sim$~4 $\times$ 10$^{-1}$ M$_{\odot}$ yr$^{-1}$. There has been a recent estimate of up to 2 $\times$ 10$^{-3}$ M$_{\odot}$ yr$^{-1}$ \citepalias{2021ApJ...910L..13E}, which is in the middle of our range. Previous studies also include that by \cite{kuo2014measuring}, who obtained an upper limit of $\sim$ 10$^{-3}$ M$_{\odot}$ yr$^{-1}$, also within our range. We note that our result is orders of magnitude below the Eddington limit,
as calculated in \citep[known as \citetalias{2019ApJ...875L...5E}]{2019ApJ...875L...5E}, meaning that M87* is in a fairly quiescent phase.

The kinetic energy per unit time of the accreting material is simply given by multiplying
half of our accretion rate by $V_A^2 \, = \, (0.23c)^2$. The minimum value of accretion energy per unit time (`thin disk', minimum $\rho$), is 6 $\times$ 10$^{33}$ J/s, or of the order of $\sim$~10$^{34}$ J/s. The maximum value of accretion energy per unit time (`thick disk', maximum $\rho$), is 6 $\times$ 10$^{37}$ J/s, or of the order of $\sim$~10$^{38}$ J/s.

We also note that the power of the jet has previously also been estimated to be $\sim$~10$^{35}$
to $\sim$~10$^{38}$ J/s \citepalias{2019ApJ...875L...5E}. The fact that our range of values is consistent with the estimated power of the jet, means that our results align with models where it is the
power of the accretion that powers the jet. 

\section*{Acknowledgments}
D.W.T. acknowledges funding support from the UK STFC through grant number ST/R000786/1.
J.S.S. acknowledges funding support from the UK STFC through grant number ST/X508329/1.
The authors have access to the EHT Collaboration through the UK membership of the JCMT.
The JCMT is operated by the East Asian Observatory on behalf of a number of international partners,
including the UK. The UK operates a consortium of Universities, who each contribute to the JCMT
running costs, along with the STFC, under the auspices of STFC grant number ST/N005856/1.

\vspace{5mm}
\facilities{EHT; JCMT}

\bibliography{bibliography}{}
\bibliographystyle{aasjournal}

\appendix
\restartappendixnumbering
\renewcommand{\theequation}{A\arabic{equation}}
\section{Calculation of Spin}
In this appendix, we provide details of the calculations for our estimates of the rotation rate of the inner region of the accretion disk, the Doppler ratio, and the spin. Note that in this paper we are seeking experimental estimates based on the EHT images of M87*, rather than exact theoretical predictions.

To estimate the spin and mass accretion rate of the supermassive black hole M87*, we assume:
\begin{itemize}
	\item  the angular size in the image provides a good estimate of the physical size of features, in particular $R_\text{ISCO}$, since the black hole is imaged close to face-on. 
	\item the accretion disk is in the equatorial plane of the black hole and is co-rotating with the spin of the black hole.
	\item  the angular momentum of the inner region of the accretion disk matches the spin of the black hole, since the spin-up timescale of a supermassive black hole is significantly shorter than its age, and furthermore we expect the spin of the black hole will be close to the upper limit of 0.998 \citep{Thorne_1974}.
	\item the intensity variation across the accretion disk is primarily due to Doppler beaming, and that, over the relatively small detection bandwidth of the EHT, the cube of the ratio of the Doppler factors for approaching and receding radiation is a Lorentz invariant \citep{Misner:1973prb}. 
\end{itemize}

We compare the calculations for three spacetime metrics: Minkowski, Schwarzchild, and Kerr, representing the flat spacetime of empty space, the spacetime of a non-rotating black hole, and the spacetime of a rotating black hole, respectively.

The Doppler factor for the Kerr metric is as follows \citep{Cisneros_2015}.
\[
D_\text{app} = \left[\frac{g_{tt}+\Omega(g_{t\phi}+\sqrt{(g_{t\phi})^2-g_{\phi\phi}g_{tt}})}{g_{tt}\sqrt{-g_{tt}-2\Omega g_{t\phi}-\Omega^2g_{\phi\phi}}}\right]^{-1}
\]
where $\Omega = \frac{\text{d}\phi}{\text{d}t} = \cos(17{\degree} )V_R/R_\text{ISCO}$ is the angular velocity of the source, and the metric components are
\[
g_{tt} = -\left(1-\frac{2mr}{\rho^2}\right)c^2\qquad g_{t\phi} =  g_{t\phi} = -\frac{2mar}{\rho^2}c\sin^2\theta
\]
\[
g_{\phi\phi} = ((r^2 + a^2)^2 - a^2\Delta \sin^2\theta)\frac{\sin^2\theta}{\rho^2}
\]
\[
g_{rr} = \frac{\rho^2}{\Delta}\qquad g_{\theta\theta} = \rho^2
\]
and where $\rho^2 = r^2 + a^2\cos^2\theta$, $\Delta = r^2 + a^2 - 2mr$, $m = GM/c^2$, and the specific spin is $a = J/Mc$ (dimensions of length) for angular momentum $J$. However, in this paper we describe spin using the dimensionless specific spin $a = Jc / GM^2$.

In the following, we set $\theta = \pi/2$ (source in the equatorial plane), and $r = R_\text{ISCO} = 5.5\times10^{13}$ m.

\subsection{Minkowski metric}
In flat Minkowski spacetime, we set $M = 0$ and $a = 0$. The Doppler factor for an approaching source then reduces to
\[D_\text{app} = [\gamma(1-\cos(\theta)\beta)]^{-1},\]
and for M87* we find the invariant Doppler ratio is
\[\left(\frac{D_\text{app}}{D_\text{rec}}\right)^3 = \left( \frac{1+0.96V_R/c}{1-0.96V_R/c}\right)^3=2.25,\]
from measurements of the EHT image in fig. 1. As shown in section 2.1 and 2.2, this gives $V_R\approx0.14c$ and hence $a \approx0.8$.

\subsection{Schwarzchild metric}
Set $a = 0$, to obtain
\[
D_\text{app} = \left[\sqrt{1-2m/r} \frac{\sqrt{1-2m/r}-\Omega r/c}{\sqrt{1-2m/r}+\Omega r/c}\right]^{-1}.
\]
With $M = 6.5\times10^9 M_\odot$, the same calculation as for Minkowski spacetime but with the cubed Doppler ratio now 2.73 yields $V_R\approx0.17c$ and $a \approx0.97$.

\subsection{Kerr metric}
Setting $a = 0.8$ and repeating the calculation using the cubed ratio of Kerr-metric Doppler factors now 2.77 results in $V_R\approx0.175c$ and $a \approx1.00$.

\subsection{Conclusion}
What we find is that, regardless of the spacetime model used, the Doppler ratio in the EHT image of M87* indicates a spin of at least 0.8, and could be close to the theoretical maximum of 0.998.\\

\section{Calculation of Accretion Rate}
\renewcommand{\theequation}{B\arabic{equation}}
\renewcommand{\thefigure}{B1}
\begin{figure*}
	\centering
	\includegraphics[width=0.8\textwidth]{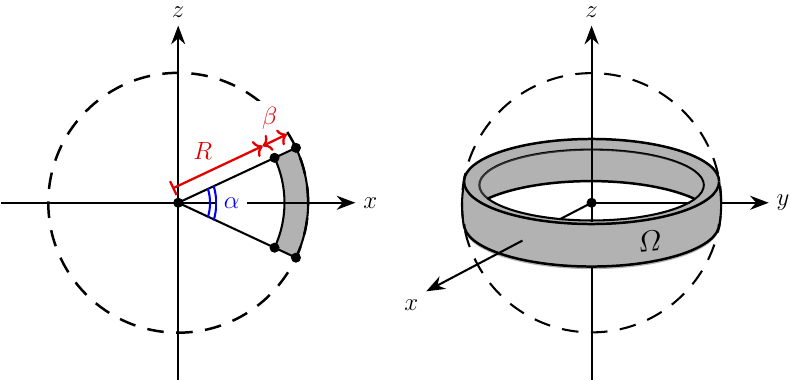}
	\caption{A schematic diagram showing the geometry of the accretion disc model, formed by an annular sector (left) revolved around the $z$-axis (right).\\[0.3cm]}
	\label{fig:GEOMDISK}
\end{figure*}

In this appendix we provide further details on calculating the accretion rates in the `thin disk' and `thick disk' approximations. We work in $\mathbb{R}^{3}$ with spherical polar coordinates $(r,\theta,\phi)\in (0,\infty)\times (0,\pi)\times (0,2\pi)$. Consider an {\it annular sector} of inner radius $R$ and thickness $\beta$, subtending an angle $\alpha$ at the centre, as shown in Figure~\ref{fig:GEOMDISK} (left). Let $\Omega$ denote the domain in $\mathbb{R}^{3}$ formed as a {\it volume of revolution} of the annular sector as it rotates around the $z$-axis, as shown on the right in Figure~\ref{fig:GEOMDISK}. The domain $\Omega$ is used to model the accreting mass:
\begin{equation*}
	\Omega \,\equiv\, (r,\theta,\phi) \text{   such that   }
    \left\{ \begin{array}{rl}
                 R\leq &r\;\leq R+\beta \\[0.4cm] 
                 \displaystyle\frac{\pi}{2}-\frac{\alpha}{2}\leq &\displaystyle\theta\;\leq\frac{\pi}{2}+\frac{\alpha}{2} \\[0.4cm] 
                 0<&\phi\;\leq 2\pi 
            \end{array}\right.
\end{equation*}
Using the spherical polar infinitesimal volume element $dV=r^{2}\sin(\theta)\,dr\,d\theta\,d\phi$, the volume of the region $\Omega$ is calculated as:
\begin{eqnarray}
	\nonumber\text{Vol}[\Omega] &=& \iiint_{\Omega}dV \\[0.2cm]
    \nonumber &=& \int_{r=R}^{R+\beta}\int_{\theta=\frac{1}{2}(\pi-\alpha)}^{\frac{1}{2}(\pi+\alpha)}\int_{\phi=0}^{2\pi} r^{2}\,\sin(\theta)\,dr\,d\theta\,d\phi \\[0.2cm] 
	\label{VolOm}	&=& 4\pi R^{2}\left( \beta + \frac{\beta^{2}}{R} + \frac{\beta^{3}}{3R^{2}}\right)\sin\left(\frac{\alpha}{2}\right).
\end{eqnarray} 
The mass of the accreting matter is given by: 
\begin{equation*}
    M = \rho\,\text{Vol}[\Omega] = 4\pi\rho\,V_{A}\, R^{2}\left( t + \frac{V_{A}\,t^{2}}{R} + \frac{V_{A}^{2}\,t^{3}}{3R^{2}}\right)\sin\left(\frac{\alpha}{2}\right)
\end{equation*}
where $\rho$ denotes the mass density and the thickness of the accreting matter $\beta=V_{A}t$ in terms of the accretion velocity $V_{A}$ and accretion time $t$. Under the assumptions of steady-state accretion and that the inner edge of the accretion disk is at the innermost stable circular orbit ($R=R_{\text{ISCO}}$), the accretion rate is then:
\begin{equation*}
    \dot{M} \,=\, \frac{dM}{dt} \,=\, 4\pi\rho\,V_{A}\,R_{\text{ISCO}}^2\left(1+\frac{V_{A}\,t}{R_{\text{ISCO}}}\right)^{\!2}\sin\left(\frac{\alpha}{2}\right).
\end{equation*}
We assume the inner radius $R$ is much larger than the thickness of the accreting matter: $R_{\text{ISCO}}\gg\beta=V_{A}t$, implying $V_{A}\,t/R_{\text{ISCO}}\ll 1$ and hence 
\begin{equation}\label{ACCRATE}
    \dot{M} \,=\, 4\pi\rho\,V_{A}\,R_{\text{ISCO}}^2\sin\left(\frac{\alpha}{2}\right).
\end{equation}
We consider two specific cases for the accretion model:
\begin{enumerate}[leftmargin=1cm,label=$\bullet$\;,itemsep=0.3cm]
	\item	{\it `Thick disk'}: we assume $\alpha\approx 60^{\circ}\approx \pi/3$ radians and hence $\sin(\alpha/2)\approx 1/2$. The accretion rate (\ref{ACCRATE}) is $\dot{M} \,=\, 2\pi\,\rho\,V_{A} R_{\text{ISCO}}^{2}$.
	\item	{\it `Thin disk'}: we assume $\alpha\approx 6^{\circ}\approx \pi/30$ radians and hence $\sin(\alpha/2)\approx \sin(\pi/60)\approx \pi/60$. The accretion rate (\ref{ACCRATE}) is then $\dot{M} \,=\, (\pi^{2}/15)\,\rho\,V_{A} R_{\text{ISCO}}^{2}$.
\end{enumerate}
These are the accretion rate approximations used for the `thick disk' and `thin disk' in the main text.

\end{document}